# Search for New Heavy Higgs Boson in B-L model at the LHC using Monte Carlo Simulation

H.M.M.Mansour, Nady Bakhet

Department of Physics, Faculty of Science, Cairo University

Email: mansourhesham@yahoo.com, nady.bakhet@cern.ch, nady.bakhet@yahoo.com

## Abstract

The aim of this work is to search for a new heavy Higgs boson in the B-L extension of the Standard Model at LHC using the data produced from simulated collisions between two protons at different center of mass energies by Monte Carlo event generator programs to find new Higgs boson signatures at the LHC. Also we study the production and decay channels for Higgs boson in this model and its interactions with the other new particles of this model namely the new neutral gauge massive boson $Z'_{B-L}$ and the new fermionic right-handed heavy neutrinos $\nu_h$.

# I. INTRODUCTION

The minimal B-L extension of the Standard Model obeying the structure $SU(3)_C \times SU(2)_L \times U(1)_L \times U(1)_{B-L}$ gauge symmetry in which the SM gauge has a further $U(1)_{B-L}$ group related to the Baryon minus Lepton (B-L) gauged number[1-2]. The minimal B-L model is defined by the condition $g'_1$ which implies no mixing between the $Z'_{B-L}$ and SM Z gauge bosons. The fields' charges are the usual SM and the B–L one where B-L = 1/3 for quarks and -1 for leptons with no distinction between generations hence ensuring universality. The B - L charge assignments of the fields as well as the introduction of a new fermionic right-handed heavy neutrinos and a scalar Higgs field with charge +2 under B−L are designed to ensure the gauge invariance.

The Standard Model is based on one complex Higgs doublet consisting of four degrees of freedom, three of which, after spontaneous Electro-Weak Symmetry Breaking turn out to be absorbed in the longitudinal polarization component of each of the three weak gauge bosons $w^\pm$ and Z. The fourth one gives the physical Higgs state. Minimal extension of the B−L model consists of a further $U(1)_{B-L}$ gauge group in addition to the SM gauge structure, three right-handed neutrinos and an additional complex Higgs singlet which is responsible for giving mass to an additional $Z'_{B-L}$ new gauge boson. Therefore the scalar sector in the B-L model consists of two real CP-even scalars that will mix together. B − L model breaking can take place at the TeV scale far below that of any Grand Unified Theory.

In the next section, we will present production cross sections, Branching Ratios and decay widths for Higgs bosons SM-like Higgs (light Higgs) $H_1$ and extra Higgs (heavy Higgs) $H_2$ by analyzing the data produced from simulated collisions between two protons at different center of mass energies by Monte Carlo event generator programs .

Also, we find the independent physical parameters of the Higgs boson in the minimal B-L extension of the Standard Model:

(1) Higgs bosons masses $M_{H1}$ , $M_{H2}$ and the scalar mixing angle α. Masses and couplings which depend on the Higgs mixing have been tested against the experimental limits obtained at the Large Electron-Positron (LEP) collider and at the Tevatron.

(2) $g'_1$ the new U (1)$_{B-L}$ gauge coupling.

(3) The mass of the new gauge boson mass $Z'_{B-L}$ . An indirect constraint on the mass of $Z'_{B-L}$ comes from the analysis at LEP :

$$\frac{M_{Z'_{B-L}}}{g'_1} \geq 7 TeV \qquad (1)$$

(4) α = 0 is the decoupling limit with $H_1$ behaving like the SM Higgs.

(5) α = $\pi/2$ is called inversion limit where $H_2$ is the SM Higgs .

| particle | $l$ | $e_R$ | $\nu_R$ | q | $\phi$ | $\chi$ |
|---|---|---|---|---|---|---|
| $Y_{B-L}$ | -1 | -1 | -1 | 1/3 | 0 | 2 |

Table 1: B − L quantum numbers for fermions and Higgs particles

The fermionic and kinetic sectors of the Lagrangian in the case of B − L extension [3] is given by:

$$L_{B-L} = i\bar{l}D_\mu\gamma^\mu l + i\bar{e}_R D_\mu\gamma^\mu e_R + i\bar{v}_R D_\mu\gamma^\mu v_R$$
$$-\frac{1}{4}W_{\mu\nu}W^{\mu\nu} - \frac{1}{4}B_{\mu\nu}B^{\mu\nu} - \frac{1}{4}C_{\mu\nu}C^{\mu\nu} \tag{2}$$

And
$$C_{\mu\nu} = \partial_\mu C_\nu - \partial_\nu C_\mu \tag{3}$$

is the field strength of the U (1)$_{B-L}$

The Higgs and Yukawa sectors of the Lagrangian are given by

$$L_{B-L} = (D^\mu\phi)(D_\mu\phi) + (D^\mu\chi)(D_\mu\chi) - V(\phi,\chi)$$
$$-(\lambda_e \bar{l}\phi e_R + \lambda_v \bar{l}\tilde{\phi} v_R + \frac{1}{2}\lambda_{v_R}\overline{v_R^c}\chi v_R + h.c.) \tag{4}$$

Where $\lambda_e, \lambda_v$ and $\lambda_{v_R}$ refer to 3 × 3 Yukawa matrices. The interaction terms $\lambda_v \bar{l}\tilde{\phi}v_R$ and $\lambda_{v_R}\overline{v_R^c}\chi v_R$ give rise to a Dirac neutrino mass term $m_D$ $\lambda_v v$ and a Majorana mass term $M_R = \lambda_{v_R} v'$ respectively. The U(1)B−L and SU (2)L × U (1)Y gauge symmetries can be spontaneously broken by a SM singlet complex scalar field χ and a complex SU(2) doublet of scalar fields $\phi$, respectively. We consider the most general Higgs potential invariant under these symmetries, which is given by:

$$V(\phi,\chi) = m_1^2\phi^+\phi + m_2^2\chi^+\chi + \lambda_1(\phi^+\phi)^2 + \lambda_2(\chi^+\chi)^2$$
$$+ \lambda_3(\chi^+\chi)(\phi^+\phi) \tag{5}$$

Where $\lambda_3 > -2\sqrt{\lambda_1 \lambda_2}$ and $\lambda_1, \lambda_2 \geq 0$ so that the potential is bounded from below. For non-vanishing vacuum expectation values (vev's) we require:

$\lambda_3^3 < 4\lambda_1\lambda_2, m_1^2 < 0$ and $m_2^2 < 0$ The vev's, $|\langle\phi\rangle| = v/\sqrt{2}$ and $|\langle\chi\rangle| = v'/\sqrt{2}$ are then given by:

$$v^2 = \frac{4\lambda_2 m_1^2 - 2\lambda_3 m_2^2}{\lambda_3^2 - 4\lambda_1\lambda_2} \tag{6}$$

That is the electroweak symmetry breaking scale $v$ and

$$v'^2 = \frac{-2(m_1^2 + \lambda_1 v^2)}{\lambda_3} \tag{7}$$

is the B − L symmetry breaking scale $v'$

After the B − L gauge symmetry breaking, the gauge field $C_\mu$ which is the new neutral massive gauge boson Z′ acquires the following mass:

$$m_{Z'_{B-L}}^2 = 4g'^2_1 v'^2 \tag{8}$$

Where g′ is the U $(1)_{B=L}$ gauge coupling constant and $v'$ is the B − L symmetry breaking scale.

The mixing between the two Higgs scalar fields, the SM complex SU(2)$_L$ doublet and complex scalar singlet is controlled by the coupling $\lambda_3$.

For positive $\lambda_3$ the B − L symmetry breaking scale $v'$ becomes much higher than the electroweak symmetry breaking scale $v$. In this case the SM singlet Higgs $\phi$ and the SM like Higgs $\chi$ are decoupled and their masses are given by

$$M_\phi = \sqrt{2\lambda_1}\,v \quad , \quad M_\chi = \sqrt{2\lambda_2}\,v' \tag{9}$$

For negative $\lambda_3$ the B−L breaking scale is at the same order of the electroweak breaking scale. In this scenario, a significant mixing between the two Higgs scalars exists and can affect the SM phenomenology. This mixing can be represented by the following mass matrix for $\varphi$ and $\chi$:

$$\frac{1}{2}M^2(\phi,\chi) = \begin{pmatrix} \lambda_1 v^2 & \dfrac{\lambda_3}{2} v v' \\ \dfrac{\lambda_3}{2} v v' & \lambda_2 v'^2 \end{pmatrix} \tag{10}$$

Therefore, the mass Eigen states fields $H_1$ and $H_2$ are given by

$$\begin{pmatrix} H_1 \\ H_2 \end{pmatrix} = \begin{pmatrix} \cos\theta & -\sin\theta \\ \sin\theta & \cos\theta \end{pmatrix} \begin{pmatrix} \phi \\ \chi \end{pmatrix} \tag{11}$$

Where the mixing angle $\theta$ is defined by

$$\tan 2\theta = \frac{|\lambda_3| v v'}{\lambda_1 v^2 - \lambda_2 v'^2} \tag{12}$$

As a result of the mixing between the two Higgs bosons, the usual couplings among the SM-like Higgs $H_1$ and the SM fermions and gauge bosons are modified. In addition, there are new couplings among the extra Higgs $H_2$ and the SM particles:

$$g_{H_1 ff} = i\frac{m_f}{v}\cos\theta \qquad g_{H_2 ff} = i\frac{m_f}{v}\sin\theta$$

$$g_{H_1 VV} = -2i\frac{m_v^2}{v}\cos\theta \qquad g_{H_2 VV} = -2i\frac{m_v^2}{v}\sin\theta$$

$$g_{H_1 z'z'} = 2i\frac{m_C^2}{v'}\sin\theta \qquad g_{H_2 z'z'} = -2i\frac{m_C^2}{v'}\cos\theta$$

$$g_{H_1 \nu_R \nu_R} = -i\frac{m_{\nu_R}^2}{v'}\sin\theta \qquad g_{H_2 \nu_R \nu_R} = i\frac{m_{\nu_R}^2}{v'}\cos\theta \qquad (13)$$

## II. RESULTS

We will present the results of simulation to discover the two Higgs bosons at LHC, the light Higgs bosons $H_1$ (SM-like Higgs) and heavy Higgs boson $H_2$ (extra Higgs) of the B-L extension of SM using Monte Carlo simulation programs Pythia8 [4-5], MadGraph5 [6], CalcHEP [7], Pythia-Calchep Interface, Root data analysis and Physical Analysis Workstation(PAW). We present these results at different energies at Large Hadrons Collider (LHC) particularly at energies in the hadronic Centre-of-Mass (CM) of 7TeV and integrated luminosity of L $\int L = 1$ fb$^{-1}$ (early discovery scenario) and at full luminosity at CM of $\sqrt{s} = 14$ TeV and an integrated luminosity of $\int L = 300$ fb$^{-1}$. First, we will present the results of production cross section for the two Higgs bosons of B-L model by two methods, SM production mechanisms method and B-L model production mechanisms. Secondly, we will present different branching ratios for both the light Higgs bosons $H_1$ and the heavy Higgs boson $H_2$ [8-9]. Also we will study some decay channels for heavy Higgs boson which depend on the scalar mixing angle $\alpha$ and finally we will present the total width for different mixing angles.

## A. Production Cross Section mechanisms

## 1. Standard production mechanisms

Figure 1 shows the normal methods for the production cross section for the SM processes as gluon-gluon fusion, vector-boson fusion W and Z, top associated production and Higgs-strahlung with scalar mixing angle zero. Figures 2 and 3 show the production cross section of the light Higgs boson H1 and the heavy Higgs boson H2 in B-L model at center of mass energy 7TeV at Large Hadrons Collider for different values of the mass . We notice that the curve is a smooth function of the mixing angle α and maximum cross section is at the smallest value of masses.

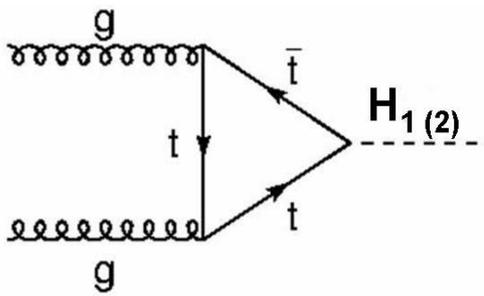

**Gluon-Gluon fusion**

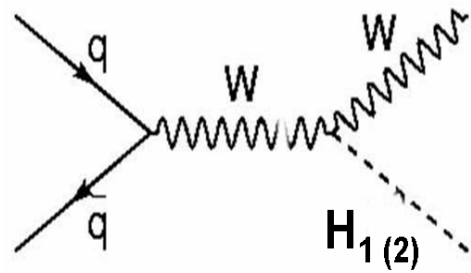

**associative production with W**

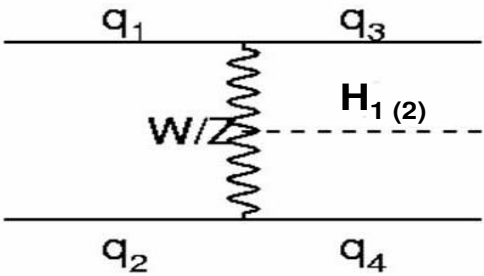

**Vector boson fusion**

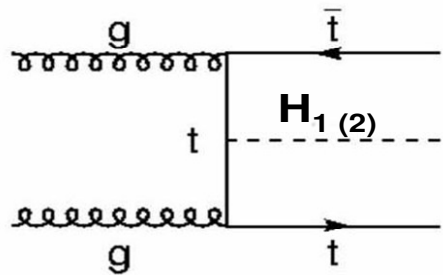

**associative production with a top pair**

**Fig. (1):** Higgs production ( light Higgs and heavy Higgs) at LHC

The cross-section for H2 at an angle α is equal to that one of H2 for π/2 − α. The maximum cross-section for H2 when α = π/2 coincides with the cross-section of H2 for α = 0. We notice that these results are in agreement with the ones that have been discussed in[2] in the minimal B − L context for high value of the mixing angle which could lead to important consequences for the Higgs boson discovery at the LHC. As in the SM, the main contribution to the production cross section comes from the gluon-gluon fusion mechanism. The next relevant contribution is given by the Higgs production in the weak vector boson mechanism.

This contribution is at the level of a few fb, as estimated above. Furthermore, the production associated with Z/W is dominant over the production associated with Z′ .we analyze the production of the heavy Higgs. It turns out that its cross sections are smaller than the light Higgs ones. The light Higgs scenario, the production associated with Z' is dominant over the Production associated with Z/W in agreement with our previous prediction.

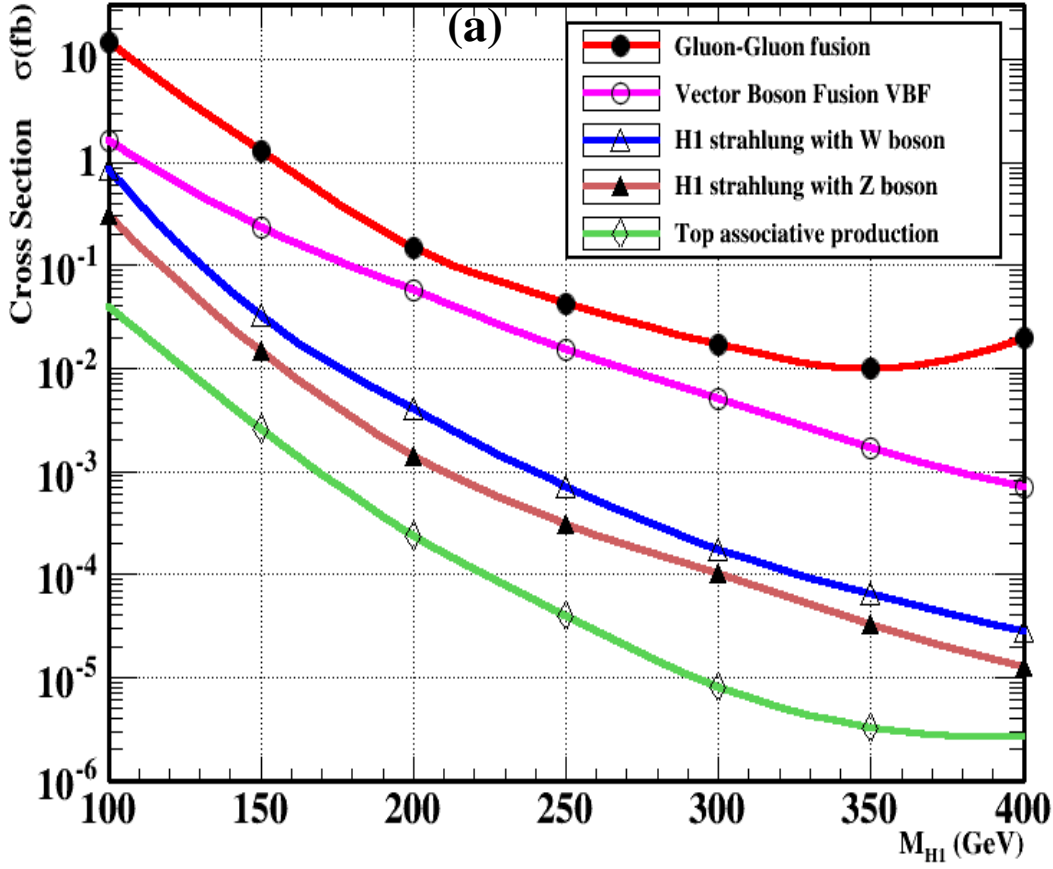

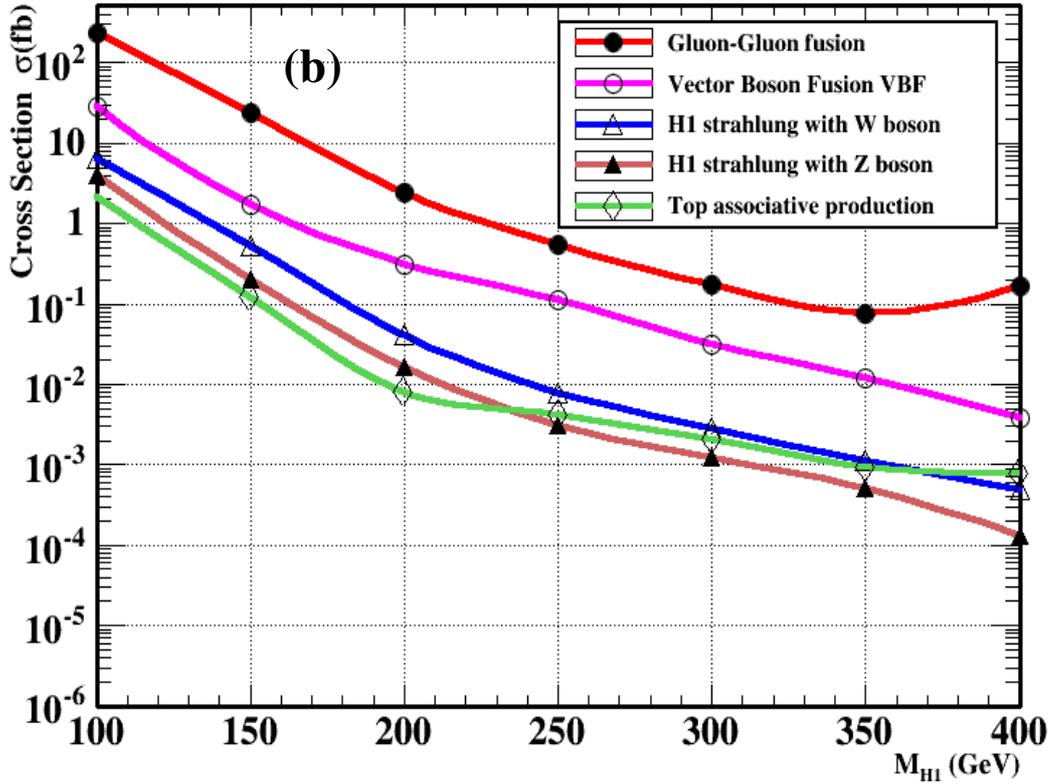

*FIG. 2: Production Cross-sections at the LHC in the B − L model for light Higgs boson $H_1$ by standard methods figure 2(a) at √s = 7 TeV and figure 2(b) at √s = 14 TeV using MadGraph5 and Pythia8 Monte Carlo event Generator*

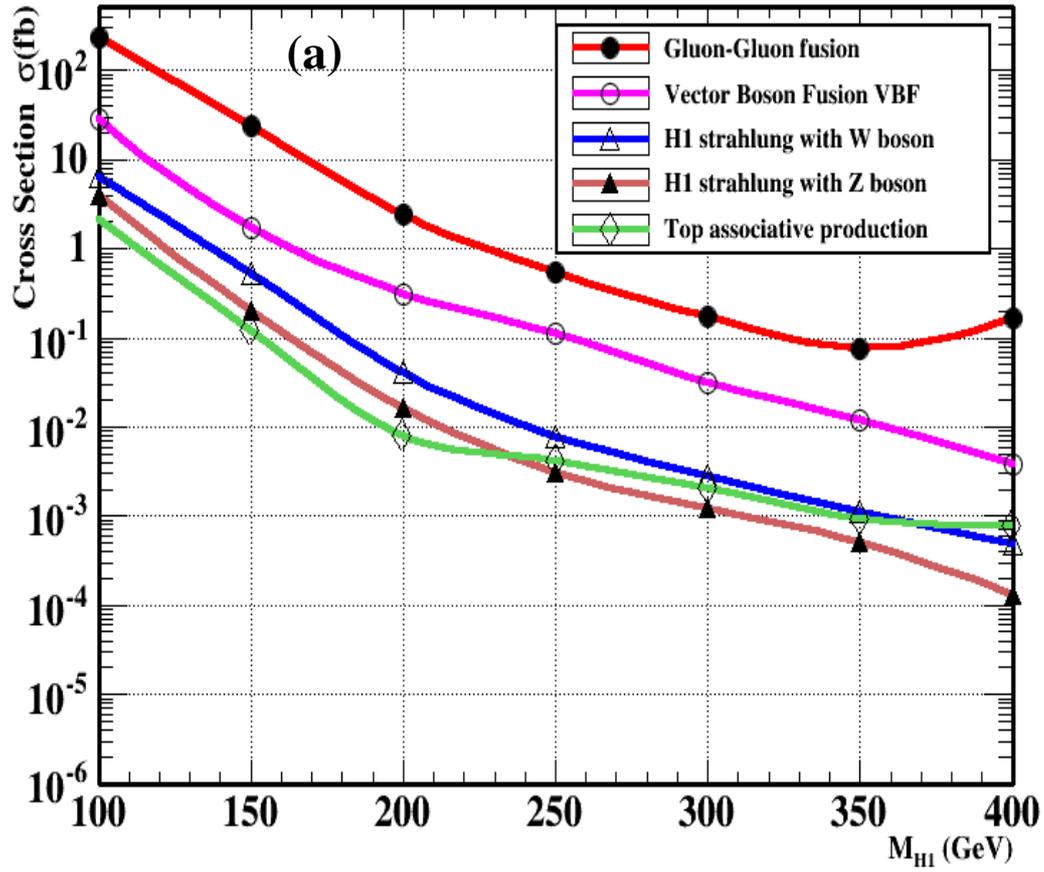

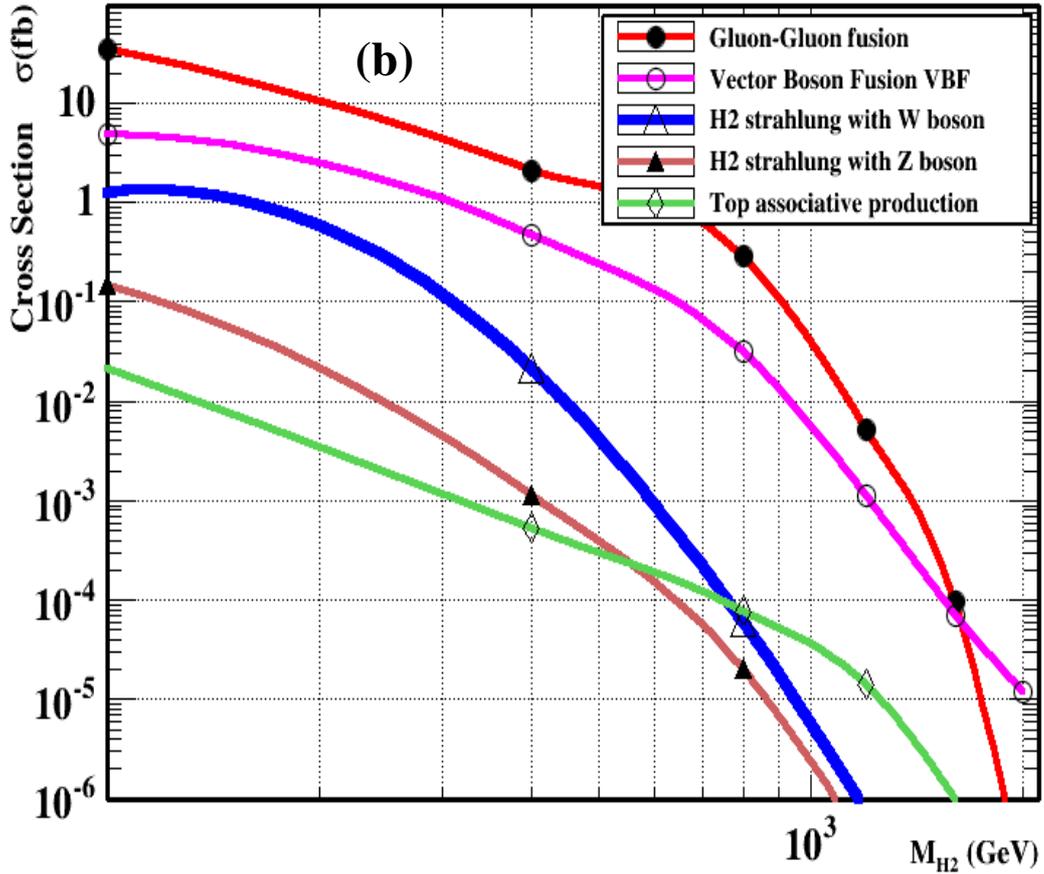

*FIG. 3: Production Cross-sections at the LHC in the B − L model for heavy Higgs boson $H_2$ by standard methods at √s = 7 TeV figure (a) and √s = 14 TeV figure (b) using MadGraph5 and Pythia8 Monte Carlo event Generator*

-١٢-

## 2. Non-Standard production mechanisms

B-L model has new particles as $Z_{B-L}$ heavy new gauge boson, three fermionic heavy neutrino right-handed [10,11,12]in addition to two Higgs bosons. These particles interact with the scalar field in the B-L model and a new mechanism will follows rather than the standard production mechanism. Figures 4 and 5 show the production cross of light and heavy Higgs by new production mechanisms at energies 7TeV and 14TeV at LHC and for different masses of the new gauge boson $Z_{B-L}$ [13,14,15] and different values for the coupling constant $g'_1$. In figure 4 the Higgs bosons light or heavy can be produced via the decay of the intermediate massive resonance gauge boson $Z_{B-L}$ according to the interaction.

$$q\bar{q} \to Z_{B-L} \to H_1 H_1 (or\ H_2 H_2) \qquad (14)$$

Figure 5 shows the decay of a heavy neutrino into a Higgs boson according to the interaction

$$q\bar{q} \to Z_{B-L} \to \nu_h \nu_h \to \nu_h \nu_l H_{1(2)} \qquad (15)$$

The interaction requires pair produced heavy neutrinos. This mechanism has the advantage that the whole decay chain can be of on-shell particles, besides the peculiar final state of a Higgs boson and a heavy neutrino. To maximize this mechanism we choose the mass of $Z_{B-L}$ = 1 TeV, $g'_1$ = 0.1 and the mass of $\nu_h$ = 200 GeV.

Figure 5(b) shows the associated production of the light Higgs boson with a photon via the SM neutral gauge bosons (γ and Z) and the new Z′ boson

$$pp \to \gamma / Z / Z' \to \gamma H_1$$

And
$$qq' \to \gamma H_1 q''q''' \qquad (16)$$

through vector-boson fusion (only W and Z bosons) for different values of the scalar mixing angle.

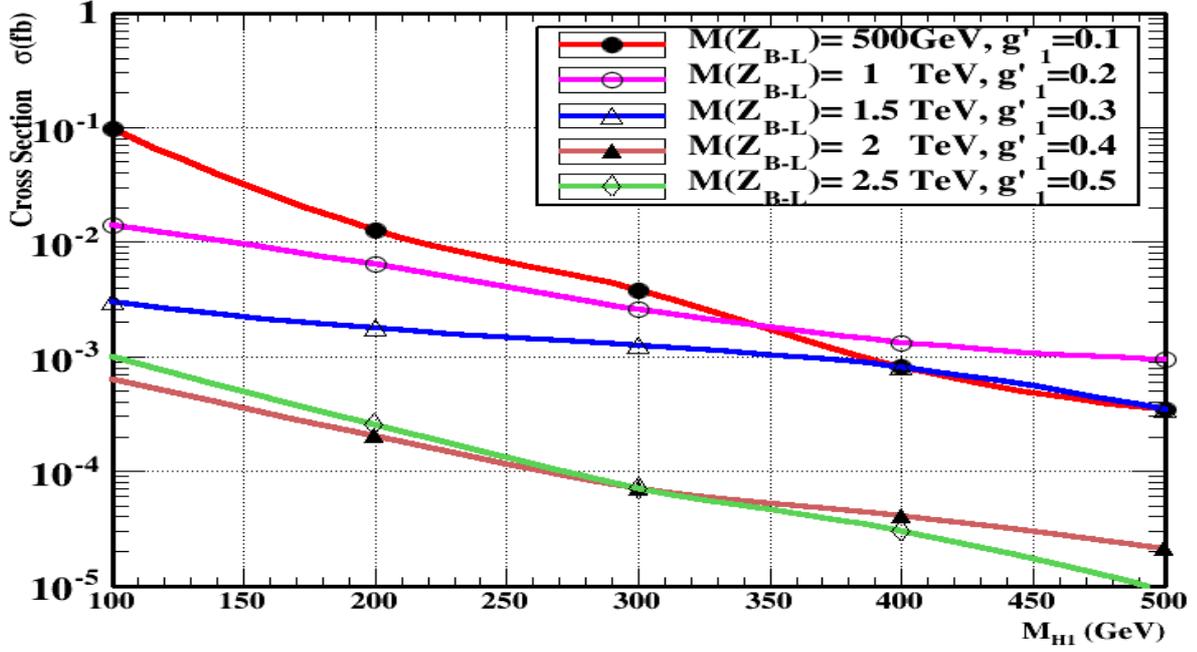

*FIG. 4(a): Production Cross-sections at the LHC in B − L model for light Higgs boson with associated production $Z_{B-L}$ gauge boson for different values of coupling constant $g'_1$ and mixing scalar angle $\alpha = \pi/3$ using MadGraph5 and Pythia8 Monte Carlo event Generator*

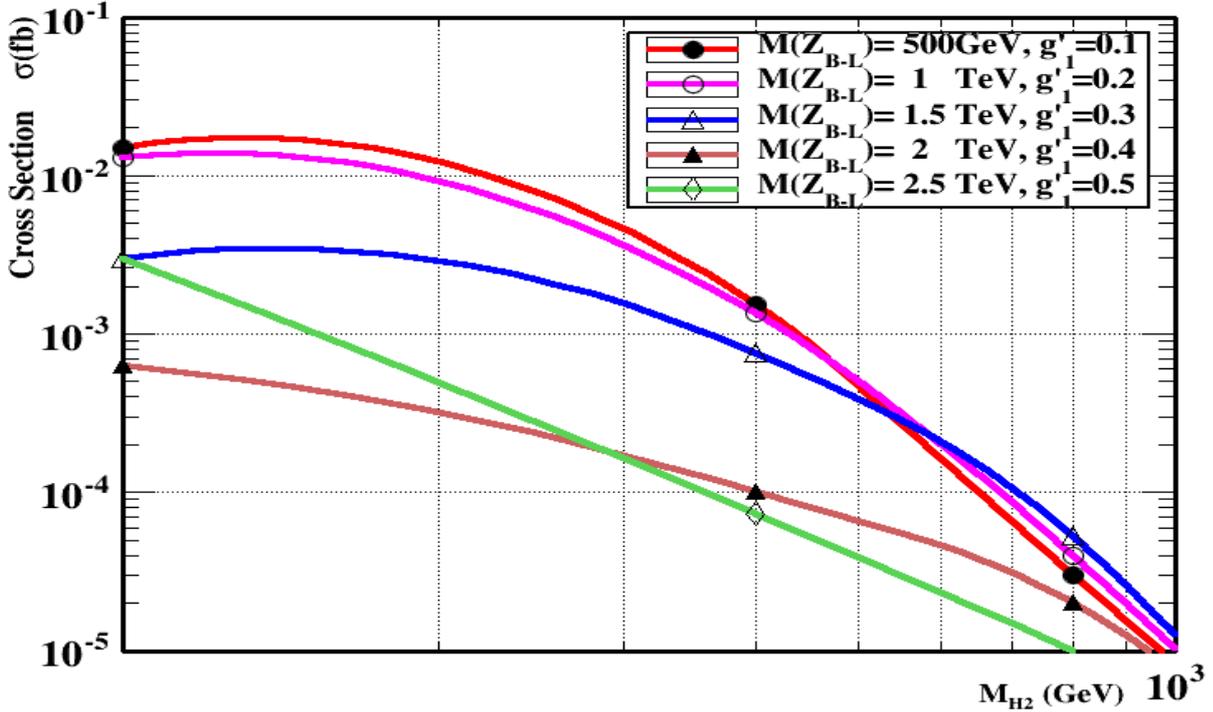

*FIG. 4(b): Production Cross-sections at the LHC in B − L model for heavy Higgs boson $H_2$ with associated production $Z_{B-L}$ gauge boson for different values of coupling constant $g'_1$ and mixing scalar angle $\alpha = 0$ using MadGraph5 and Pythia8 Monte Carlo event Generator*

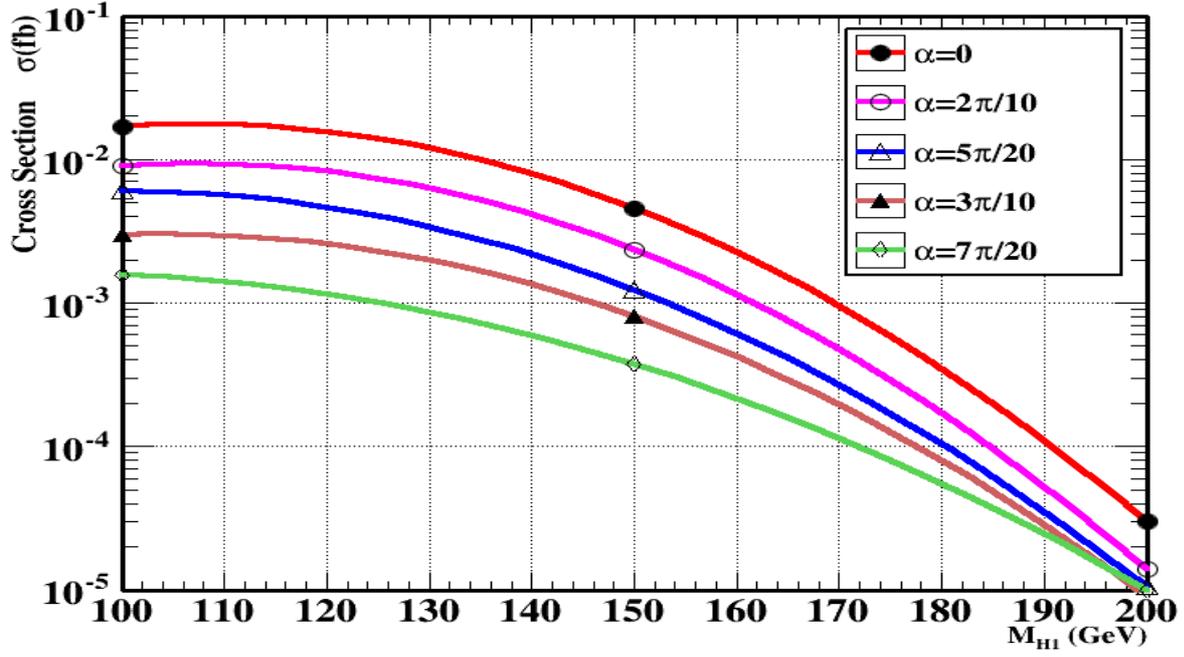

*FIG. 5(a): Production Cross-sections at the LHC in B − L model for the associated production of light higgs $H_1$ with one heavy and one light neutrinos for different values of scalar mixing angle by using MadGraph5*

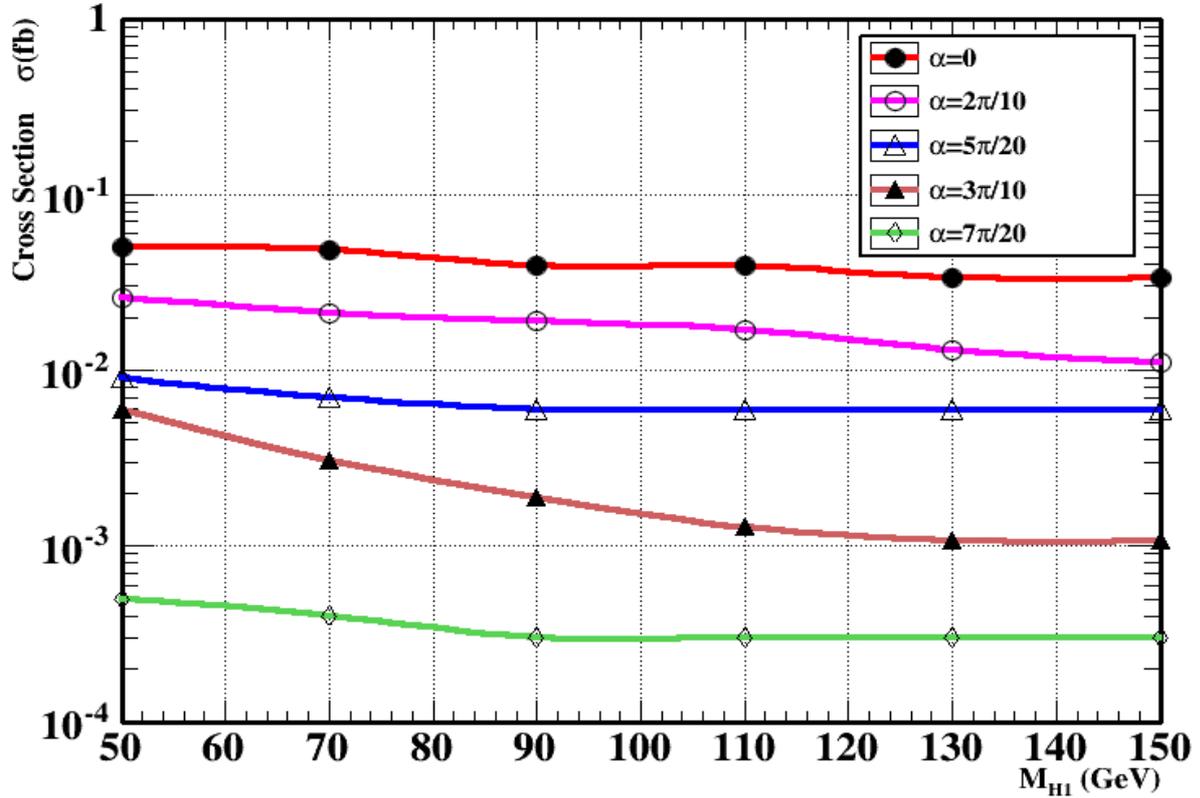

*FIG. 5(b): Production Cross-sections of light higgs $H_1$ at the LHC in B − L model in the vector-boson fusion for different values of scalar mixing angle by using MadGraph5*

## B. Branching ratios

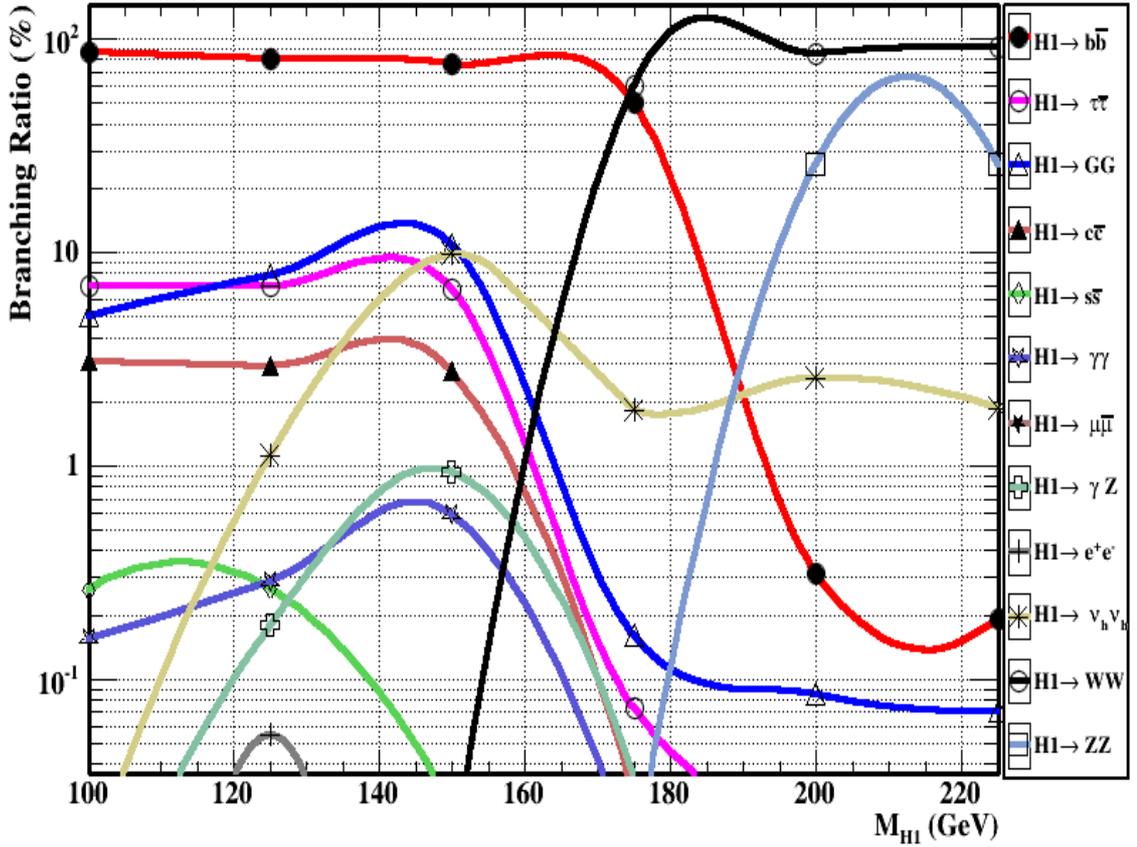

*FIG. 6(a): Branching ratios of light Higgs $H_1$ for mixing angle α = 5π/20 and the mass of heavy neutrino =60 GeV .*

Now we will discuss all two body decay channels of Higgs boson and branching ratios for each channel.

Figure 6(a) gives all possible decay channels of light Higgs bosons by using MadGraph5 and the Pythia8 event generator. The light Higgs boson has 12 decay channels. It has a new decay channel which is the heavy neutrino channel and we assume here the mass of heavy neutrino to be 50GeV. Also we used a fixed scalar mixing angle angle α = 5π/20 . We note that the decay channel of light Higgs to bottom-antibottom quarks initially has the highest branching ratio value at small value of heavy neutrino mass. We also note that when the mass of light Higgs begins to increase the decay channel of the light Higgs to pair of charged vector bosons W becomes the dominant branching ratio. Another decay channel which follows that is, The SM neutral vector boson decay channel. So from any of these channels one can get 4 leptons in the final state plus missing energy from ZZ decay or WW decay. In this case we can detect the light Higgs at LHC through these channels. Also in the case of the B − L model the decay of light Higgs into two heavy neutrino pairs is a very important signature besides being an interesting feature of the B − L model at LHC.

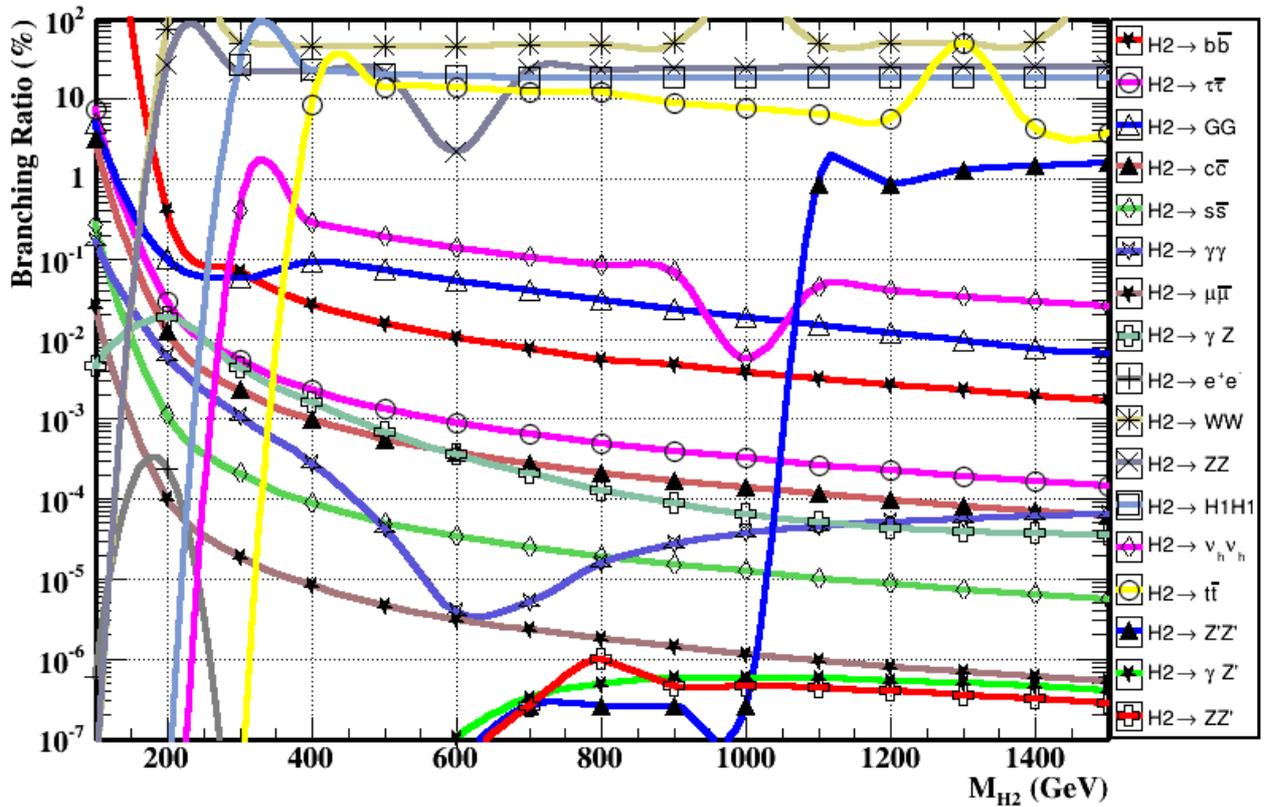

*FIG. 6(b): Branching ratios of heavy Higgs $H_2$ for mixing angle α = π/6 and the mass of heavy neutrino =150 GeV*

Figure 6(b) shows the branching ratios of all two body decay channels of the heavy Higgs boson $H_2$ for a range of mass 100 GeV to 1500 GeV. In this case we have a new decay channel for the heavy Higgs in B-L model where it can decay into a pair of new Z'$_{B-L}$ boson and this channel is not in the SM. one can consider this as a new feature of the B-L model at LHC where the new gauge boson can decay to a pair of heavy neutrino and every one of them can decay into 2 leptons.

The other standard decay channels of the heavy Higgs decaying into W boson pairs is always dominant when it is kinematically open. Before that the decay into bottom quarks is a dominant decay channel, and the Higgs boson decays into pairs of photons.

The decay branching ratios of the light H1 and heavy Higgs H2 bosons are shown in Figures 6(a) and 6(b), respectively, as functions of the Higgs masses. As expected, the branching ratio of the light Higgs are very close to the SM ones.

Now, we discus the new three decay channels of three heavy Higgs boson in the B-L model which are H$_2$ → H$_1$H$_1$ (decay to light Higgs), H$_2$ → Z'$_{B-L}$ Z'$_{B-L}$ (decay to new gauge boson ) and H$_2$ → V$_h$ V$_h$ (decay to heavy neutrinos) see figure 7

The partial decay width to pair of the lighter Higgs channel is:

$$\Gamma(H_2 \to H_1 H_1) \approx \frac{1}{16\pi\sqrt{2}} \frac{g^2_{H_1^2 H_2}}{m_{H_2}} (1 - \frac{4m^2_{H_2}}{m^2_H})^{1/2} \qquad (17)$$

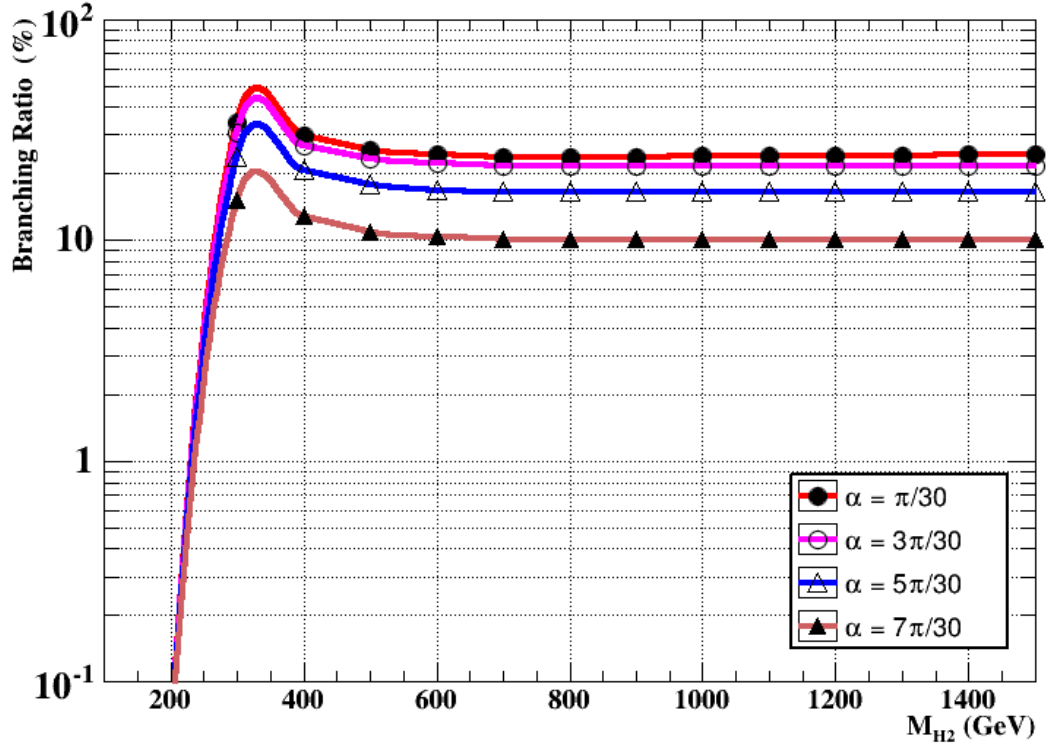

***FIG. 6(c): Branching ratio of decay channel $H_2 \rightarrow H_1 H_1$ for different mixing angle values***

Figure 6(c) shows the decay channel of heavy Higgs boson into pairs of the light Higgs boson $H_2 \rightarrow H_1 H_1$ see figure 7 in the B − L model for different values of scalar mixing angle for different values of H2 mass. The contribution of this process to the total width is not negligible when the mixing angle is small, this channel vanishes when α tends to zero.

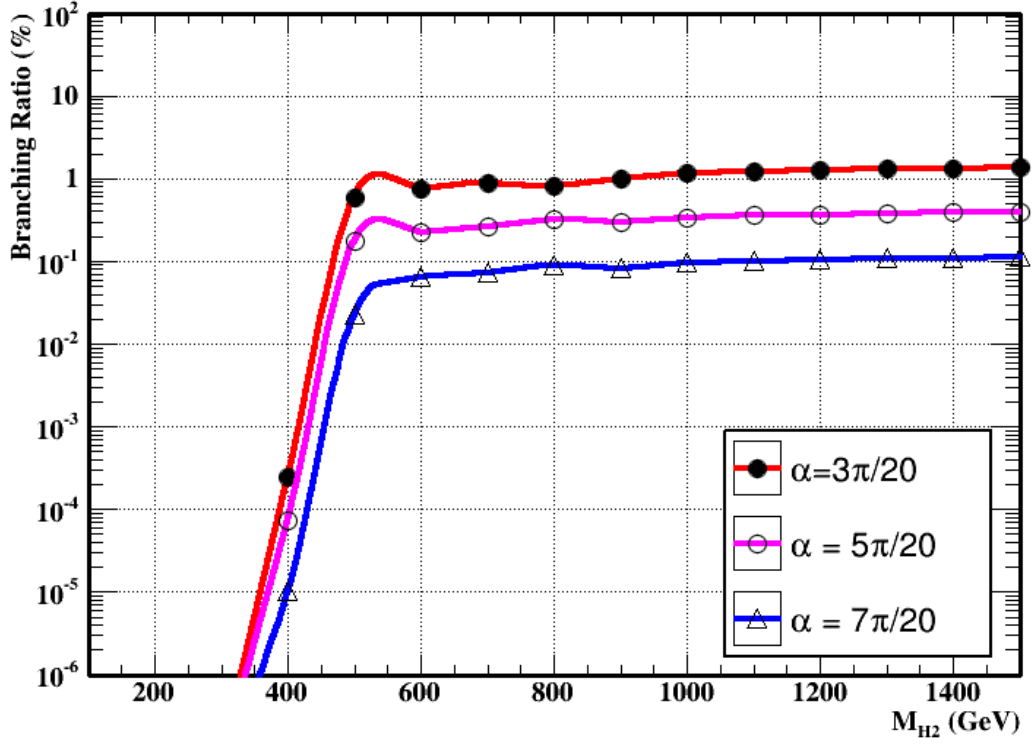

***FIG. 6(d): Branching ratio of decay channel $H_2 \rightarrow Z_{B-L}Z_{B-L}$ for different mixing angle values***

Figure 6(d) shows the decay channel of heavy Higgs boson into pairs of the new neutral massive gauge boson $H_2 \rightarrow Z_{B-L}Z_{B-L}$ see figure 7 in the B − L model for different values of the scalar mixing angle and the branching ratios for this channel at different masses of the heavy Higgs . From the figure, the dependence on the mixing angle α of the branching ratios of H2 into pairs of non-SM particles ,the interaction of the heavy Higgs boson with SM (or non-SM) particles has an overall sin α (or cos α respectively) dependence. Nonetheless, the branching ratios in the figure depend also on the total width , for α > π/4 is dominant by the H2 $\rightarrow$ W$^+$W$^-$ decay. Hence, when the angle assumes big values ,the angle dependence of the H2 branching ratios into heavy neutrino pairs and into Z′ boson pairs follow a simple cot α behavior.

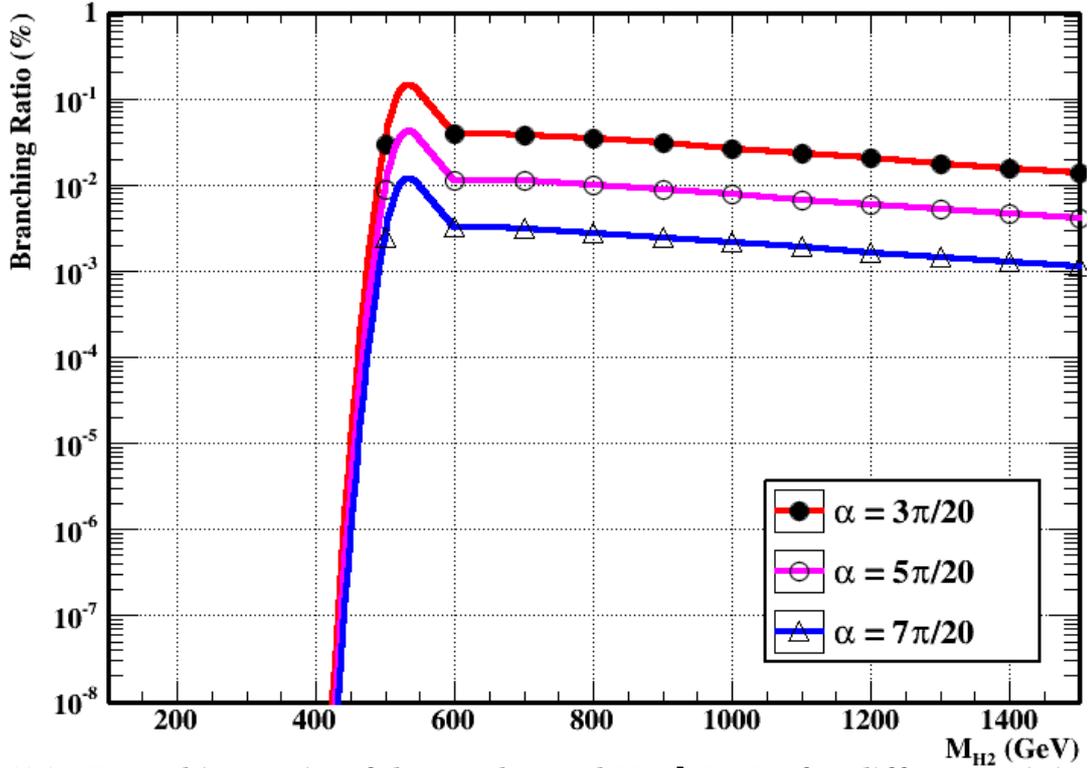

*FIG. 6(e): Branching ratio of decay channel $H_2 \rightarrow V_h V_h$ for different mixing angle values*

Figure 6(e) shows the third new decay channel of the heavy Higgs boson into pairs of heavy neutrinos $H_2 \rightarrow V_h V_h$ in the B − L model for different values of the scalar mixing angle and the branching ratios for this channel at different masses of the heavy Higgs .

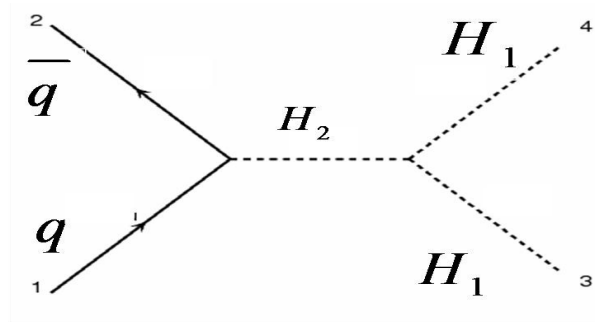

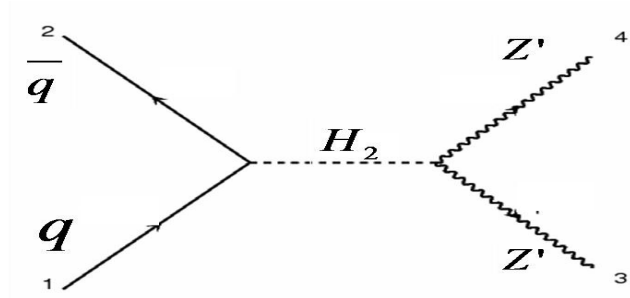

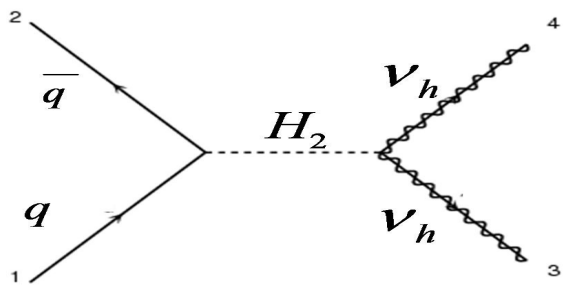

*FIG.(7)Three decay channels of heavy Higgs H2 to three non-SM particles, light Higgs H1, new gauge boson Z' and heavy neutrino $V_h$*

## B. Decay Width

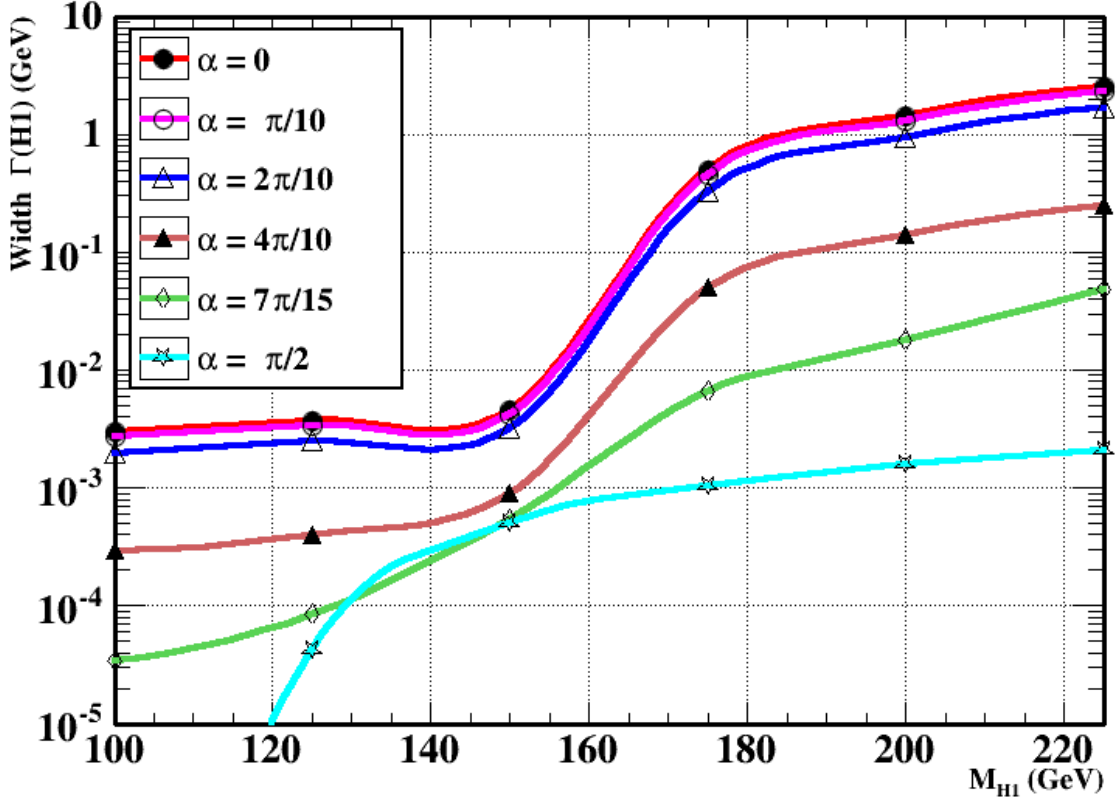

*FIG. 8(a): Total width of light Higgs $H_1$ for different mixing angle α and the mass of heavy neutrino =75 GeV by using MadGraph5 program*

The Higgs particle tends to decay into the heaviest gauge bosons and fermions allowed by the phase space. The Higgs decay modes can be classified into three categories: Higgs decays into fermions, Higgs decays into massive gauge bosons, and Higgs decays into mass less gauge bosons. The decay widths into fermions are directly proportional to the H➔f f couplings:

$$\Gamma(H_1 \to ff) \approx m_{H_1}(\tfrac{m_f}{v})^2(1-\tfrac{4m_f^2}{m_{H_1}^2})^{3/2}\cos^2\theta$$

$$\Gamma(H_2 \to ff) \approx m_{H_2}(\tfrac{m_f}{v})^2(1-\tfrac{4m_f^2}{m_{H_2}^2})^{3/2}\sin^2\theta \tag{18}$$

the decay widths into massive gauge bosons V = Z′, Z, W are directly Proportional to the HVV couplings. This includes two-body, three-body, and four-body decays

For V=W boson or Z boson:

$$\Gamma(H_1) \approx \tfrac{m_{H_1}^3}{v^2}(1-\tfrac{4m_V^2}{m_{H_1}^2})^{3/2}\cos^2\theta$$

$$\Gamma(H_2) \approx \tfrac{m_{H_2}^3}{v^2}(1-\tfrac{4m_V^2}{m_{H_2}^2})^{3/2}\sin^2\theta$$

For V= Z′$_{B-L}$ new boson:

$$\Gamma(H_1) \propto \tfrac{m_{H_1}^3}{v'^2}(1-\tfrac{4m_V^2}{m_{H_1}^2})^{3/2}\sin^2\theta$$

$$\Gamma(H_2) \propto \tfrac{m_{H_2}^3}{v'^2}(1-\tfrac{4m_V^2}{m_{H_2}^2})^{3/2}\cos^2\theta \tag{19}$$

The mass less gauge bosons are not directly coupled to the Higgs bosons, but they are coupled via W, charged fermions, and quark loops. This implies that the decay widths are in turn proportional to the HVV and Hff couplings, hence they are relatively suppressed.

Figures 8(a) and 8(b) show the total widths for $H_1$ and $H_2$, respectively for different values of the scalar mixing angle . We note from figure 8(a) that the total width increase through mass range from 100 GeV to 225 GeV of $H_1$ mass. Initially, with small mass of the light Higgs the width is small until the contribution of the W width to the total width of $H_1$ which increase it. from the above Equations, one finds that all decay widths of the light Higgs are proportional to $\cos 2\theta$, except the new decay mode of $Z'$. Furthermore, this channel has a very small contribution to the total decay width. Therefore, the light Higgs branching ratios (the ratios between the partial decay widths and the total decay width) have small dependence on the mixing parameter $\theta$. Thus, it is expected to see no significant difference between the results of the light Higgs branching ratios in this model of $B - L$ extension and the SM ones. On the other hand, the heavy Higgs branching ratios have relevant dependence on $\theta$.

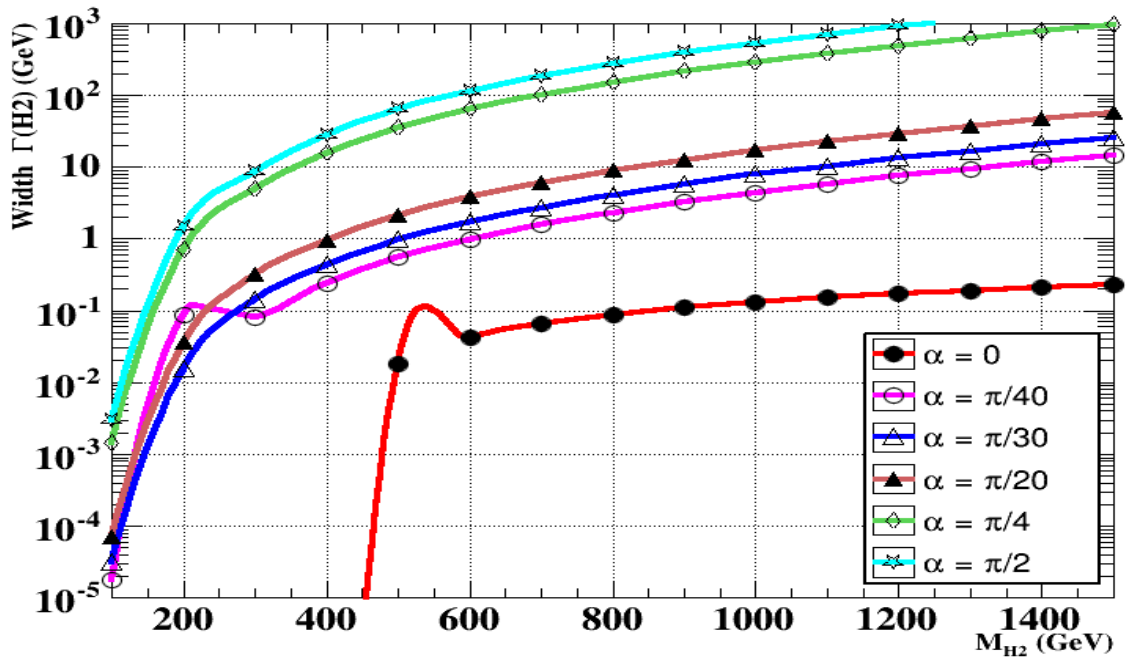

*FIG. 8(b): Total width of heavy Higgs $H_2$ for different mixing angle α by using MadGraph5*

## III. CONCLUSIONS

We have simulated the production and decay of the two Higgs states in B-L model at LHC at different energies using MC programs MadGraph 5, Pythia8 and CalcHEP. We calculated the production cross section of light and heavy Higgs by using normal methods of Standard Model and new methods of B-L model and we found that the heavy Higgs has relatively small cross sections but it is accessible at LHC, also we presented all possible decay channels for light and heavy Higgs states and their branching ratios, the total width for each state and focused on new decay channels of heavy Higgs boson in the B-L model into a pair of heavy neutrino or pair of new gauge boson where the new heavy Higgs can be detected at LHC using one of these channels.

## ACKNOWLEDGEMENTS

It is a pleasure to thank L. Basso (Freiburg Univ. Germany) for useful discussions of B-L model, T. Sjostrand (Lund Univ. Sweden) for useful discussions of PYTHIA, J. Alwall (Stanford Univ. USA) for useful discussions of Mad-Graph5/MadEvent and C. Duhr also many thanks to the administration of faculty of science Cairo university Egypt particular to Prof. Gamal Abd El Nasir, prof. Sheriff Mourad and Prof. Omar Osman also many thanks to Prof. M.I. Wanas the director of CTP-BUE Egypt and Prof. Amr Radi the director of ENHEP-ASRT Egypt.